\documentclass{iopjournal}
\usepackage{amsmath,amssymb,amsfonts}
\usepackage{enumitem}
\usepackage{comment}
\usepackage[
backend=biber,
style=ieee,
maxnames=3,
minnames=1,
doi=false,
isbn=false,
url=false,
eprint=false,
date=year
]{biblatex}
\newcommand{\mydim}[1]{\mathbb{R}^{#1}}
\newcommand{\dreyer}{\texttt{Dreyer2023}}
\newcommand{\lee}{\texttt{Lee2019}}

\begin{document}

\articletype{Paper} 

\title{Quantifying Event-Related (De)Synchronization Variability for Brain-Computer Interface: A Unified and Interpretable Framework}

\author{Simon Kojima$^{1,*}$\orcid{0000-0002-6509-7363} and Fabien Lotte$^{1}$\orcid{0000-0002-6888-9198}}

\affil{$^1$Inria Centre at the University of Bordeaux, Talence, France}

\affil{$^*$Author to whom any correspondence should be addressed.}

\email{simon.kojima@inria.fr}

\keywords{Brain-Computer Interface, Motor Imagery, Electroencephalography, Variability, Metrics}

\begin{abstract}
Objective:
Brain-Computer Interfaces (BCIs) enable the control of external devices by decoding user intentions from electroencephalography (EEG).
However, substantial EEG variability within and between users remains a major challenge.
To better understand this variability, we propose interpretable metrics that independently quantify temporal, spatial, and frequency variability in BCI-related brain activity within and between users.
Approach:
We propose a framework to quantify variability by extracting EEG features and defining variability as their dispersion around their centroid using appropriate distance functions. Using two motor imagery BCI datasets ($N = 133$ users), we investigated the relationship between BCI performance and the variability metrics through within-user and cross-user classification experiments.
Main results:
Negative correlations of $-0.2$ to $-0.4$ were observed across most conditions, suggesting that lower variability is associated with higher BCI performance. Moreover, the metrics revealed differences in robustness to variability between the deep learning and Riemannian-based classifiers, with the former showing weaker correlations.
The results demonstrate the effectiveness of the proposed variability metrics and suggest that reducing variability may improve BCI performance while revealing differences in the sensitivity of classification models to different types of variability.
Significance:
The framework quantifies temporal, spatial, and frequency variability at multiple hierarchical levels (within-trial, between-trial, and between-trial-group), providing interpretable measures to better understand EEG variability and support more robust BCIs. It could also be used to characterize dataset variability, evaluate classifier sensitivity, incorporate variability into objective functions, and provide variability-based user feedback.
\end{abstract}

\section{Introduction}
\label{sec:introduction}

Brain-Computer Interfaces (BCIs) infer user intentions from brain activity to control external devices without movement~\cite{wolpaw_Braincomputerinterface_2000b}.
Motor Imagery BCIs (MI-BCIs) decode imagined movements from electroencephalography (EEG) signals~\cite{padfield_EEGBasedBrainComputer_2019} by detecting event-related (de)synchronization (ERD/S) when users imagine moving specific body parts.
ERD/S is a decrease (ERD) or increase (ERS) in EEG power, mainly in the mu ($8$--$13$\,Hz) and beta ($13$--$30$\,Hz) bands.

However, some users cannot operate MI-BCIs~\cite{vidaurre_CureBCI_2010a}, a phenomenon known as BCI inefficiency~\cite{hammer_Psychologicalpredictors_2012}.
Even among those who can, many show low control accuracy, making BCI use unreliable.
One of the major factors contributing to BCI inefficiency and unreliability, is the variability in responses across users, as well as within-user variability caused by the nonstationarity of EEG signals~\cite{saha_IntraIntersubject_2020a, lotte_eusipco2026}.
Several variability factors related to MI-BCI performance have been identified~\cite{zhang_Subjectinefficiency_2020, vongroll_Largescale_2024b, jeunet2016advances}.
However, the mechanisms underlying the variability of responses in MI-BCIs remain insufficiently understood.

On the other hand, many Signal Processing (SP) and Machine Learning (ML) approaches have been proposed to address BCI variability~\cite{lotte_reviewclassification_2018, lotte_eusipco2026}.
For example, Transfer Learning (TL) approaches have been used to mitigate variability across sessions (session-to-session TL) or users (user-to-user TL)~\cite{wu_TransferLearning_2022}.
This TL approach included domain adaptation to handle variability by aligning or adjusting the distribution of responses~\cite{rodrigues_RiemannianProcrustes_2019, he_TransferLearning_2020}.
Moreover, attempts have been made to reduce variability through user training~\cite{zhang_Subjectinefficiency_2020}. 

While the approaches above may have reduced the impact of variabilities to some extent, they are still far from having eradicated the problem altogether.
In order to build variability-robust BCIs, it is first necessary to deeply understand the characteristics of these variabilities. 
Without such understanding, it is difficult to propose dedicated and effective SP \& ML methods, or targeted user training strategies.
In turn, to understand variability, we first need to measure and quantify such variabilities.
However, studies focusing on the quantitative characterization of variability remain relatively limited.

Previous studies have proposed methods to quantify between-trial stability within each class in MI-BCI tasks by representing the responses of each class as covariance matrices and computing their dispersion on a Riemannian manifold~\cite{lotte_Definingquantifying_2018}. 
In addition, other studies have quantified between-trial variability by computing the standard deviation across trials of the mean ERD/S amplitude for each trial~\cite{rimbert_EventRelatedDesynchronization_2022}.
Ivanov et al. proposed a method that divides each trial into sub-epochs, computes a covariance matrix for each, and then clusters these matrices. The trajectories of each trial in the subspace spanned by the cluster centroids are analyzed, and their variability serves as a metric called the multi-class intra-trial trajectory (MITT)~\cite{ivanov_MultiClassIntraTrial_2026}.
One study quantified between-trial variability per EEG channel as the inverse of the mean Pearson correlation across all trial pairs~\cite{demelo_procedureminimize_2024a}.
Another assessed between-subject variability using Spearman correlations between subject-wise averaged ERD/S patterns~\cite{wriessnegger_InterIntraindividual_2020}.

However, these methods cannot independently quantify variability in the three key BCI signal modalities: temporal, spatial, and frequential.
Moreover, Variability in BCI arises at multiple hierarchical levels.
First, it can be broadly divided into between-user variability and within-user variability.
Between-user variability refers to differences in EEG responses across users, even when the same task is performed using the same system.
It is a major factor degrading performance in cross-user classification, where a classifier trained on one group is applied to another.
It also explains why most current online BCIs rely on within-user classifiers.
Within-user variability, on the other hand, refers to the variability observed within the same user.
This can be further divided into two levels: within-trial variability and between-trial variability.
Within-trial variability represents, e.g., how much the response fluctuates within a single trial when a user sustains a task for several seconds. 
In contrast, between-trial variability refers to the variability in responses across repeated executions of the same task by the same user.
Existing methods are generally designed to quantify specific types of variability.
For example, MITT measures within- and between-trial spatial covariance variability, STDERD measures between-trial temporal variability, and class stability measures between-trial spatial covariance variability.
As a result, they cannot separately quantify variability at each hierarchical level.

In this study, we propose a method that can independently measure variability across the three modalities mentioned above (temporal, spatial, and frequential), while providing high interpretability and enabling the evaluation of variability at multiple hierarchical levels, here within-trial, between-trial, and between-trial-group. 
The between-trial-group level can be used, for example, to quantify variability between users or to evaluate variability between sessions of the same user.
Furthermore, in the proposed framework, features are extracted for each modality at each hierarchical level, and variability is defined as the dispersion of these features around their mean. 
This formulation results in a set of highly interpretable metrics. 
In addition, the definition of the mean used for computing variability, as well as the distance measure between the mean and each data point, can be flexibly chosen, allowing the method to adapt to various data characteristics.

Moreover, the framework has been released as an open-source Python package \texttt{Nearby} with high compatibility with the widely used MNE-Python ecosystem, making it readily accessible to a broad range of researchers: \href{https://simonkojima.github.io/nearby-docs/}{https://simonkojima.github.io/nearby-docs/}.


The following Section~\ref{sec:metrics} presents the mathematical formulation of the proposed variability metrics.
The parameter settings, feature representations, and distance functions used in the experimental evaluation are described in Section~\ref{sec:eval_metrics}.

\section{Newly Proposed Variability Metrics}
\label{sec:metrics}

We propose new metrics to quantify neural response variability in EEG-based BCIs. Although applicable to any ERD/S paradigm, we focus on motor imagery BCIs and analyze the variability of ERD/S patterns elicited by an MI task.

First, we define ERD/S as follows:
\begin{equation}
\label{eq:erds}
X[i,c,f,t]
=
\frac{P[i,c,f,t]-P_{BL}[i,c,f]}
{P_{BL}[i,c,f]}
\times 100,
\end{equation}
Here, $P$ and $P_{BL}$ denote the time--frequency representations of power during the task interval and the baseline interval of each trial, respectively~\cite{pfurtscheller_EventrelatedEEG_1999}. $i$, $c$, $f$, and $t$ denote the trial, EEG channel, frequency bin, and time sample, respectively, and $P_{BL}[i,c,f]$ denotes the baseline power averaged over the baseline time window for trial $i$, channel $c$, and frequency bin $f$.
$X$ is a 4-dimensional tensor and expressed as percentage changes relative to the baseline power.

Following ERD/S computation, observations are defined by selecting subsets of trials, channels, frequency bins, and time samples relevant to the variability metric of interest.
Specifically, let \(\mathcal{I}\), \(\mathcal{C}\), \(\mathcal{F}\), and \(\mathcal{T}\) denote the sets of selected trials, channels, frequency bins, and time samples, respectively.
The proposed variability metrics are computed independently for each class, and these sets may differ across classes depending on the corresponding data.
For example, when quantifying the variability of mu-rhythm ERD/S during left-hand motor imagery, $\mathcal{I}$, $\mathcal{C}$, $\mathcal{F}$, and $\mathcal{T}$ would correspond to the left-hand MI trials, 8--13 Hz frequency range, task interval, and channels associated with left-hand motor imagery responses (e.g., C4 and neighboring channels).

Here, we define three operators to extract ERD/S slices in the spatial ($\phi_s$), frequency ($\phi_f$), and temporal ($\phi_t$) domains.
For each operator, the data are averaged over all dimensions except the target modality.
Let $Z \in \mathbb{R}^{C\times F\times T}$ be an ERD/S pattern, where $C$, $F$, $T$ denote the number of EEG channels, frequency bins, and time samples, respectively.

\begin{equation}
\label{eq:phi_t}
\phi_t(Z)
=
\frac{1}{CF}
\sum_{c=1}^{C}
\sum_{f=1}^{F}
Z[c,f,t]
\end{equation}

\begin{equation}
\label{eq:phi_s}
\phi_s(Z)
=
\frac{1}{FT}
\sum_{f=1}^{F}
\sum_{t=1}^{T}
Z[c,f,t]
\end{equation}

\begin{equation}
\label{eq:phi_f}
\phi_f(Z)
=
\frac{1}{CT}
\sum_{c=1}^{C}
\sum_{t=1}^{T}
Z[c,f,t]
\end{equation}

In this study, EEG variability is quantified as the average distance between ERD/S observations $x_i$ and a representative reference $M$ (typically an average across trials).
The observations $x_i$ are obtained by applying one of the operators $\phi_t$, $\phi_s$, or $\phi_f$ to an ERD/S tensor, depending on the modality of interest.
For example, a temporal observation is extracted as
$
x_i=\phi_t(X[i,\mathcal C,\mathcal F,\mathcal T]).
$
The variability is then computed as
\begin{equation}
\label{eq:dispersion}
v = \frac{1}{n}\sum_{i=1}^{n}\mathrm{distance}(x_i,M),
\end{equation}
where $n$ denotes the number of observations (i.e., sub-epochs, trials, or trial groups), and $M$ is the representative reference computed from the corresponding observations.

\subsubsection{Within-Trial Variability}

Within-trial variability quantifies the extent to which the ERD/S pattern changes within a single trial.
Specifically, each trial is divided into multiple sub-epochs, and the variability across sub-epochs is evaluated using a slice of ERD/S extracted from each sub-epoch.

First, a ERD/S pattern of trial $j$ ($X[j, \mathcal C, \mathcal F, \mathcal T],\, j \in \mathcal I$) is segmented into $n$ sub-epochs of length $L$\,s with stride $\Delta$\,s, yielding $X_e \in \mydim{n \times |\mathcal C| \times |\mathcal F| \times T_e}$.
Here, $T_e$ denotes the number of time samples in each sub-epoch corresponding to the sub-epoch length $L$.
Within-trial variability is computed using~\eqref{eq:dispersion}, where the distance function and the definition of the representative reference $M$ are determined according to the type of variability to be evaluated.
For the within-trial variability metrics, the observation index $i$ in \eqref{eq:dispersion} corresponds to the sub-epoch index $(i = 1,\ldots,n)$.

In this study, within-trial temporal variability mainly captures changes in amplitude over time, whereas spatial and frequency variability aim to evaluate the stability of the frequential and spatial pattern across time.
Accordingly, Euclidean distance is used for temporal features, while angular distance is used for spatial and frequential features.

\paragraph{WiTrialTemp (Within-Trial Temporal) Variability}
For temporal variability, a sub-epoched ERD/S slice corresponding to the temporal modality is extracted.
Specifically, the channels and frequency bands of interest in $X_e$ are averaged to obtain $x_i=\phi_t(X_e[i])\in\mydim{T_e}$, which represents the temporal waveform within the $i$-th sub-epoch.

The variability is then computed using the squared Euclidean distance and~\eqref{eq:dispersion}.
The distance function is defined as:
$
\mathrm{distance}(x_i, M) = (\bar{x}_i - M)^2,\;
\bar{x}_i = \frac{1}{T_e} \sum_{t=1}^{T_e} x_i(t),
$
where the reference value $M$ is given as the Euclidean mean:
$M = \frac{1}{n}\sum_{i=1}^{n} \bar{x_i}$.
This metric quantifies how much the ERD/S amplitude varies across sub-epochs within a trial.

\paragraph{WiTrialSpat (Within-Trial Spatial) Variability}
For spatial variability, each sub-epoch is represented by its mean topographic pattern.
ERD/S is averaged over frequency and time of interest within each sub-epoch to obtain $x_i = \phi_c(X_e[i]) \in \mydim{|\mathcal C|}$, which corresponds to the mean spatial distribution of ERD/S in sub-epoch $i$.

To evaluate changes in spatial patterns without being affected by amplitude scaling, we use an angular distance:
\begin{equation}
\label{eq:angular-distance}
\mathrm{distance}(x_i,\,M) =
\arccos\left( \frac{x_i \cdot M}{\|x_i\|_2\,\|M\|_2} \right).
\end{equation}
Angular distance is defined as the angle between two vectors and can evaluate scale-invariant differences in pattern shape.
The representative reference $M$ is defined as the Fr\'{e}chet mean on a unit hypersphere:
\begin{equation}
\label{eq:angular-frechet-mean}
M = \operatorname*{argmin}_{\|m\|_2=1}
\sum_{i=1}^{n}
\arccos\left( \frac{x_i \cdot m}{\|x_i\|_2\,\|m\|_2} \right).
\end{equation}
In this paper, this optimization was performed using the Riemannian steepest descent algorithm implemented in Pymanopt~\cite{pymanopt}.
This metric captures how much the ERD/S topography (spatial pattern) varies across sub-epochs.

\paragraph{WiTrialFreq (Within-Trial Frequency) Variability}
For frequency variability, each sub-epoch is represented by its mean spectral pattern.
ERD/S is averaged over channel and time within each sub-epoch to obtain $x_i =\phi_f(X_e[i]) \in \mydim{|\mathcal F|}$, i.e., the mean frequency spectrum of sub-epoch $i$.

As in spatial variability, angular distance is used to focus on changes in frequential pattern while ignoring amplitude scaling.
Therefore, distance computation and estimation of the representative reference $M$ are performed using~\eqref{eq:angular-distance} and~\eqref{eq:angular-frechet-mean}, respectively.
This metric quantifies how much the ERD/S spectral pattern varies across sub-epochs within a trial.

\subsubsection{Between-Trial Variability}
Between-trial variability quantifies the extent to which ERD/S patterns change across multiple trials.
Let the time--frequency representation of ERD/S for trials of interest be $X_{\mathrm{trials}} = X[\mathcal{I}, \mathcal{C}, \mathcal{F}, \mathcal{T}]$.
For the between-trial variability metrics, the observation index $i$ in \eqref{eq:dispersion} corresponds to the trial index $i \in \mathcal I$, and the number of observations is given by $n = |\mathcal I|$.

In this study, for the temporal modality, we used Dynamic Time Warping (DTW), which can evaluate waveform-shape similarity while accounting for temporal alignment across trials~\cite{sakoe_Dynamicprogramming_1978}.
In contrast, for the spatial and frequency modalities, we used angular distance to evaluate changes in pattern shape independently of amplitude scaling.

\paragraph{BtwTrialTemp (Between-Trial Temporal) Variability}
For the temporal modality, the ERD/S time series of each trial is used.
ERD/S is averaged over the channel and frequency axes within each trial to obtain $x_i = \phi_t(X_{\mathrm{trials}}[i]) \in \mydim{|\mathcal T|}$.

Inter-trial distances are computed using DTW. DTW finds an optimal nonlinear alignment between two time series and provides a distance that reflects differences in waveform~\cite{sakoe_Dynamicprogramming_1978}.
The representative reference $M$ is defined as the Euclidean mean of $x_i$ across trials.
This metric quantifies the variability of ERD/S waveform patterns across trials.

\paragraph{BtwTrialSpat (Between-Trial Spatial) Variability}
For the spatial modality, each trial is represented by its mean topographic pattern.
ERD/S is averaged over frequency and time within each trial to obtain $x_i = \phi_s(X_{\mathrm{trials}}[i]) \in \mydim{|\mathcal C|}$.

As in WiTrialSpat, angular distance is used.
Therefore, distance computation and estimation of the representative reference $M$ are performed using~\eqref{eq:angular-distance} and~\eqref{eq:angular-frechet-mean} across trials, respectively.
This metric captures the variability of ERD/S topographic (spatial) patterns across trials.

\paragraph{BtwTrialFreq (Between-Trial Frequency) Variability}
For the frequency modality, each trial is represented by its mean spectral pattern.
ERD/S is averaged over the channel and time axes within each trial to obtain $x_i = \phi_f(X_{\mathrm{trials}}[i]) \in \mydim{|\mathcal F|}$.

As in the spatial modality, angular distance is used, and the distance computation and estimation of the representative reference $M$ are performed using~\eqref{eq:angular-distance} and~\eqref{eq:angular-frechet-mean} across trials, respectively.
This metric quantifies the variability of ERD/S spectral patterns across trials.

\begin{table}[t]
\centering
\caption{Summary of the proposed variability metrics.}
\label{tab:variability-metrics}
\resizebox{\linewidth}{!}{
\begin{tabular}{l|l|l|l|l|l}
\hline
\textbf{Metric} & \textbf{Level}      & \textbf{Extractred Slice of ERD/S $x_i$}   & \textbf{Modality} & \textbf{Distance} & \textbf{Reference $M$} \\
\hline \hline

WiTrialTemp     & Within-trial        & ERD/S timecourse per sub-epoch             & Temporal    & $(\bar x_i - M)^2$   & Euclidean mean of $\bar x_i$ \\
WiTrialSpat     & Within-trial        & Mean topography per sub-epoch              & Spatial     & Angular distance     & Fr\'echet mean \\
WiTrialFreq     & Within-trial        & Mean spectrum per sub-epoch                & frequential & Angular distance     & Fr\'echet mean \\
\hline
BtwTrialTemp    & Between-trial       & ERD/S time series per trial                & Temporal    & DTW                  & Euclidean mean \\
BtwTrialSpat    & Between-trial       & Mean topography per trial                  & Spatial     & Angular distance     & Fr\'echet mean \\
BtwTrialFreq    & Between-trial       & Mean spectrum per trial                    & frequential & Angular distance     & Fr\'echet mean \\
\hline
BtwTrialGrpTemp & Between-trial-group & Representative ERD/S time series per group & Temporal    & DTW                  & Euclidean mean \\
BtwTrialGrpSpat & Between-trial-group & Representative topography per group        & Spatial     & Angular distance     & Fr\'echet mean \\
BtwTrialGrpFreq & Between-trial-group & Representative spectrum per group          & frequential & Angular distance     & Fr\'echet mean \\
\hline
\end{tabular}
}
\end{table}

\subsubsection{Between-Trial-Group Variability}
\label{sec:metrics_BtwTrialGrp}

This metric quantifies variability between trial groups.
By appropriately defining trial groups, the same formulation can be used to assess, for example, between-session or between-user variability.

Let $\mathcal I_k \subseteq \mathcal I$ denotes the set of trial indices from group $k$.
The group $k$ representative ERD/S pattern is computed as the Euclidean mean over trials:
\begin{equation}
\label{eq:trial_group}
\bar X_k
=
\frac{1}{|\mathcal I_k|}
\sum_{i\in\mathcal I_k}
X[i,\mathcal C,\mathcal F,\mathcal T],\,\,
\bar X_k
\in
\mydim{|\mathcal C|\times|\mathcal F|\times|\mathcal T|}
\end{equation}

Then, $\bar X_k$ are stacked along the group axis to form
$X_{\mathrm{tg}} \in \mydim{K\times|\mathcal C|\times|\mathcal F|\times|\mathcal T|}$,
with the first dimension indexing the groups.

From each group $k$, an ERD/S slice is extracted from $X_{\mathrm{tg}}[k, \mathcal C, 
\mathcal F, \mathcal T]$ using $\phi_t$, $\phi_s$, and $\phi_f$, for temporal, spatial, and frequential metrics, respectively.
The extracted slices are then used as the observations $x_i$ in \eqref{eq:dispersion} to compute between-trial-group variability across groups.
The observation index $i$ in \eqref{eq:dispersion} corresponds to the group index $k$, and the number of observations is given by $n = K$.
The distance function and the representative reference $M$ are chosen according to the evaluation target.
In this study, DTW was used for the temporal modality, whereas angular distance was used for the spatial and frequency modalities.

The representative reference $M$ can be defined flexibly.
Typically, an average (e.g., Euclidean mean, or Fr\'{e}chet mean depending on the distance measure) is used; however, it is also possible to fix $M$ to a reference pattern corresponding to a specific subject or session.
This enables quantification of dispersion relative to the ERD/S pattern of, e.g., a particular user or session.
Between-trial-group variability includes three variants:
BtwTrialGrpTemp (Between-Trial-Group Temporal Variability),
BtwTrialGrpSpat (Between-Trial-Group Spatial Variability),
and BtwTrialGrpFreq (Between-Trial-Group Frequency Variability).
The distance measures are the same as those used for between-trial variability:
DTW is used for BtwTrialGrpTemp, and angular distance is used for BtwTrialGrpSpat and BtwTrialGrpFreq.

\subsubsection{Summary}
In summary, we proposed a set of metrics to quantify EEG variability based on the unified dispersion formulation in~\eqref{eq:dispersion}.
Table~\ref{tab:variability-metrics} summarizes the extracted ERD/S slices, distance measures, and centroid (reference) estimation methods used in each metric.
The proposed framework enables independent evaluation of variability in temporal, spatial, and frequential modalities under a unified formulation.

Moreover, this unified expression provides a basic definition of dispersion, and in particular, it reduces with the definition of variance when using the squared Euclidean distance.
In addition, the distance measures adopted in this study are all widely used, and the contents of variability captured by each metric (temporal waveform, spatial, and frequential pattern) are intuitive to interpret.
Therefore, the proposed variability metrics are considered to have high interpretability.

\section{Evaluation}
\label{sec:methods}
In this study, we compute the above variability metrics on two large data sets, then conduct two classification experiments (within-user and cross-user), to evaluate the metrics relationship with BCI classification performance. We also study two different classifiers (a representative Riemannian geometry one and a deep learning one), to explore which variability modalities best explain their performance.

\subsection{Dataset Description}

We used two open datasets, (A) \dreyer{}~\cite{dreyer_largeEEG_2023} and (B) \lee{}~\cite{lee_EEGdataset_2019}, which were accessed via MOABB~\cite{chevallier_largestEEGbased_2024}.
For both datasets, only EEG recordings were used.
The original publications report that informed consent was obtained from all participants and that the studies received approval from the respective local ethics committees.

\vspace{3pt}
(A) \dreyer{}: 
It contains EEG recordings from 87 users performing left- and right-hand MI tasks~\cite{dreyer_largeEEG_2023}.
For each user, a single-session EEG recording consisting of six runs was provided.
The first and second runs constituted the acquisition phase, during which data were collected for training the classifier, whereas runs 3 to 6 comprised the online phase, in which users received real-time feedback from the BCI.
In each run, 20 trials were performed per class, i.e., left- and right-hand MI.
Among the 87 users, five (4, 29, 41, 59, and 86) were excluded from further analysis due to excessive noise in the recordings (4, 29, 41, and 86) and missing runs (59), leaving 82 users for analysis.
27 EEG channels were used.

\vspace{3pt}
(B) \lee{}: 
It contains EEG recordings from 54 users performing left- and right-hand MI tasks~\cite{lee_EEGdataset_2019}.
For each user, two-session EEG recordings consisting of two runs (a training run and a test run) of MI tasks were provided.
As the test run does not include true class labels (see the original paper for details~\cite{lee_EEGdataset_2019}), only the training run was used in this study.
Each run consisted of 100 trials (50 trials per class).
Among the 54 users, three (12, 34, and 47) were excluded due to excessive noise in the recordings, leaving 51 users for analysis.
Sixty-two EEG channels were used.

\subsection{Variability Metrics}
\label{sec:eval_metrics}
In this study, we evaluated the nine metrics proposed in Section~\ref{sec:metrics} together with four existing metrics.
The four existing metrics were STDERD~\cite{rimbert_EventRelatedDesynchronization_2022}, class\_stability~\cite{lotte_Definingquantifying_2018}, MITT\_InterTrialVar, and MITT\_IntraTrialVar~\cite{ivanov_MultiClassIntraTrial_2026}.
STDERD quantifies how much the mean ERD amplitude varies across trials, for each class. 
The class\_stability metric measures how stable each trial (represented as a covariance matrix) is on the Riemannian manifold (i.e., it represents the $1 / (1 + \mathrm{variability})$, such that larger class\_stability values indicate lower variability). 
MITT is an approach based on clustering trial covariance matrices and quantifying variability by analyzing inter- and intra-trial dispersion of temporal trajectories along the feature dimensions defined by cluster centroids.

As the datasets used consist of left- and right-hand MI tasks, metrics were computed separately for each class (left / right), and then averaged across classes.
Specifically, for the temporal modality, the metrics were calculated using C4 and C3 for the left-hand and right-hand classes, respectively. For the spatial modality, all channels were used. For the frequency modality, the metrics were computed for each class using both C3 and C4, and the resulting four variability values (2 classes $\times$ 2 channels) were averaged.
For the temporal dimension, the MI task duration was extracted, and for the frequency dimension, the $8$--$30$\,Hz range was used.

These metrics were computed at three hierarchical levels: within-user, cross-user (subject-wise), and cross-user (group-wise).
First, in the within-user analysis, a total of 10 variability metrics were computed for each user, including the proposed within-trial and between-trial metrics, as well as existing metrics (class\_stability, STDERD, and two MITT metrics).

Next, in the cross-user analysis, two types of variability metrics were evaluated: subject-wise and group-wise.
For the subject-wise analysis, the within-user variability metrics were first computed for each subject and then averaged across the subjects in each user group (i.e., the training or test user).
Let $S$ denote the set of subjects and let $v(s)$ denote the variability metric computed for subject $s \in S$. The subject-wise variability is defined as:
$\frac{1}{|S|}\sum_{s\in S} v(s).$
For the group-wise analysis, a representative ERD/S pattern was first computed for each subject using \eqref{eq:trial_group}. The representative patterns were then stacked to form
$X_{\mathrm{tg}} \in \mydim{|S| \times |\mathcal C| \times |\mathcal F| \times |\mathcal T|}$,
and ERD/S slices were extracted according to the target modality, and the variability was computed as described in Section~\ref{sec:metrics_BtwTrialGrp}.

We additionally evaluated a test-user-referenced variant of the group-wise analysis.
In this variant, the extracted slice of the test subject, $x^{s_t}$, was used as the reference $M$, whereas the slices from the remaining subjects,
$\{x^{s}\mid s\in S,\; s\neq s_t\}$,
were used as the observations in \eqref{eq:dispersion}.
These metrics are denoted by the suffix ``-TR'', for example, BtwTrialGrpTemp-TR.
Note that the cross-user (group-wise) analysis was only applicable to the proposed metrics (BtwTrialGrp variability); therefore, existing metrics were not included in this analysis.

As EEG preprocessing for computing the metrics, a fourth-order Butterworth filter was applied bidirectionally in the $1$--$45$\,Hz band.
Epochs were extracted relative to the task onset, from $0.5$ to $5.0$\,s for \dreyer{} and from $0.5$ to $4.0$\,s for \lee{} , and then downsampled to $128$\,Hz.

To estimate the metrics' ERD/S patterns, time-frequency representations were computed using the Discrete Prolate Spheroidal Sequences (DPSS) multitaper method~\cite{slepian_Prolatespheroidal_1978} (time-bandwidth product = 4, number of tapers = 3, window length = 1\,s) with a frequency resolution of 1\,Hz.
ERD/S was computed using \eqref{eq:erds} with a $-2$ to $0$,s baseline, and the resulting time courses were decimated by a factor of 2.
For STDERD, the mu and beta bands ($8$--$30$\,Hz) were extracted and averaged across the frequency dimension.
Since class\_stability and the MITT-based metrics are covariance-matrix-based (and not ERD/S-based), the epoch data after band-pass filtering ($8$--$30$\,Hz)  and downsampling ($128$\,Hz) were used.
Existing metrics were computed following the same procedures as in the original papers (class\_stability:~\cite{lotte_Definingquantifying_2018}, MITT-based metrics:~\cite{ivanov_MultiClassIntraTrial_2026}, STDERD:~\cite{rimbert_EventRelatedDesynchronization_2022}).

For the WiTrial-based metrics, sub-epochs were generated by segmenting the task interval into 0.1\,s sub-epochs with a 0.1\,s stride.
For the within-user analysis, since \lee{} contains data from two sessions, metrics were computed separately for each session and then averaged across sessions.

\subsection{Classification}
In this study, classification was performed under two conditions: within-user and cross-user.
For within-user classification, a classifier was trained and evaluated using data from the same user.

In all cases, the following preprocessing was applied to the data.
First, a fourth-order Butterworth band-pass filter was applied bidirectionally within the range of 8--30\,Hz.
Next, for \dreyer{}, epochs were extracted from 0.5--5.0\,s relative to the task onset, while for \lee{}, epochs were extracted from 0.5--4.0\,s.
All data were then resampled to 128\,Hz.

For classification, both a Riemannian-based classifier, TSLR (tangent space + logistic regression)~\cite{yger_RiemannianApproaches_2017b}, and DeepConvNet~\cite{schirrmeister_Deeplearning_2017} were used.
However, only TSLR was used for within-user classification because the limited amount of training data available for each user is generally insufficient for training deep learning models effectively~\cite{chevallier_largestEEGbased_2024}.
In TSLR, covariance matrices with Ledoit--Wolf shrinkage~\cite{ledoit_wellconditionedestimator_2004} estimated from the epoched data were used as features.
For logistic regression, the lbfgs solver with L2 regularization ($C=1.0$) and a maximum of $10000$ iterations was used.
The TSLR implementation was based on pyriemann~\cite{pyriemann}.

For DeepConvNet training, Cross Entropy Loss was used together with a cosine annealing learning rate schedule ($T_{\mathrm{max}}=500$, $\eta_{\min}=1\times10^{-6}$).
The AdamW optimizer was used with a learning rate of $5\times10^{-4}$ and a weight decay of $1\times10^{-2}$.
Training was performed for up to 500 epochs with a batch size of 64.
Early stopping was applied if the validation loss did not improve for 75 consecutive epochs.
The model with the lowest validation loss was used for evaluation.
The DeepConvNet implementation was based on Braindecode~\cite{aristimunha_Braindecodetoolbox_2025}.

Training was conducted with a fixed random seed.
The data were normalized for each channel to have zero mean and unit variance.
The normalization parameters were estimated from the training data and applied to the validation and test data.

\subsection{Within-user classification}
In the within-user classification, TSLR models were trained using data from each user, and evaluated on unseen trials from the same user and session.

For \dreyer{}, runs 1 and 2 were used for training (40 trials per class in total), while runs 3 to 6 were used for testing (80 trials per class).

For \lee{}, two sessions are available.
Classification accuracy was computed separately for each session, and the average was reported as the final accuracy.
Specifically, for each class, the first 40 trials were used for training and the remaining 10 trials were used for testing.

\subsection{Cross-user classification}
In the cross-user classification scenario, the same pipeline was applied to both datasets.
Data from 20 users were used for training, and 1 user was used for testing.
The classification procedure was as follows:

\begin{enumerate}[itemsep=0pt]
    \item From the full set of users, 20 users were randomly selected for training and 1 user for testing.
    For DeepConvNet, the 20 selected training users were further split into 16 training users and 4 validation users.
    \item All the aforementioned preprocessing steps were applied.
    \item Domain adaptation was applied (details below).   
    \item Classification was performed using TSLR and DeepConvNet.
    \item Steps 1--4 were repeated 100 times.
\end{enumerate}

For domain adaptation, Riemannian Procrustes Analysis (RPA)~\cite{rodrigues_RiemannianProcrustes_2019} was applied for TSLR, and Euclidean Alignment (EA)~\cite{he_TransferLearning_2020} was applied for DeepConvNet.
The adaptation was performed separately for each session of each user.
For RPA, only recentering and rescaling were applied.
For validation and test users, domain adaptation was applied in a sequential manner.
Specifically, to simulate a realistic BCI usage scenario, the transformation for the $N$-th trial was computed using only the data from the first $N$ trials.

For \dreyer{}, in TSLR, the numbers of training and test trials were 2400 and 120, respectively.
In DeepConvNet, the numbers of training, validation, and test trials were 1920, 480, and 120, respectively.
In contrast, for \lee{}, in TSLR, the numbers of training and test trials were 2000 and 100, respectively.
In DeepConvNet, the numbers of training, validation, and test trials were 1600, 400, and 100, respectively.

\subsection{Statistical Analysis}
The correlations between the obtained variability scores and classification performance were computed to evaluate the relationship between changes in variability and classification performance. All analyses described below were conducted independently for each dataset.
Spearman correlation coefficients were computed between the variability metrics and the classification performance.
The resulting $p$-values were corrected for multiple comparisons using the Benjamini--Hochberg (BH) procedure~\cite{benjamini_ControllingFalse_1995}.
Specifically, 10 correlations were computed for the within-user classification analysis, 20 ($2$ models $\times$ $10$ metrics) for the cross-user (subject-wise) analysis, and 12 ($2$ models $\times$ $6$ metrics) for the cross-user (group-wise) analysis.
For all obtained correlation coefficients, the standard errors (SE) were estimated using a bootstrap method with 10000 resamples.

\begin{figure*}[!t]
\centerline{\includegraphics[width=\linewidth]{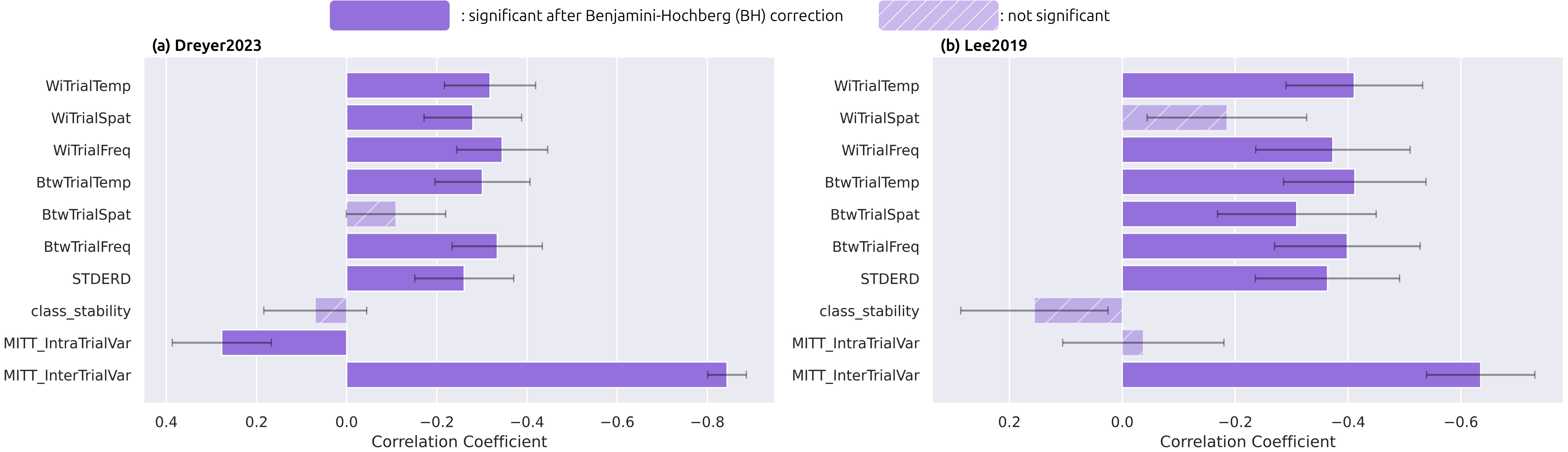}}
\caption{
Spearman correlation coefficients between each variability metric and classification performance in within-user classification. (a) \dreyer{} and (b) \lee{}. Solid colors indicate results that were significant after Benjamini-Hochberg multiple comparison correction, while hatched patterns indicate results with no significant correlation. Error bars represent the standard errors.
}
\label{fig:within-user}
\end{figure*}

\section{Results}
\label{sec:results}
Fig.~\ref{fig:within-user} shows the results for the within-user classification.
In (a) \dreyer{}, significant correlations were observed for all metrics except BtwTrialSpat and class\_stability.
Furthermore, all metrics showed negative correlations except for MITT\_IntraTrialVar.
In (b) \lee{}, significant negative correlations were observed for all metrics except WiTrialSpat, class\_stability, and MITT\_IntraTrialVar.
Notably, MITT\_InterTrialVar showed a strong negative correlation of approximately $-0.6$ to $-0.8$ in both datasets.

Fig.~\ref{fig:cross-user_subject-wise} shows the results for the test group in the cross-user (subject-wise) analysis.
The results for the training user group in the cross-user (subject-wise) analysis are shown in Fig.~1 of the Supplementary material.
Interestingly, no significant correlations were observed between the variability metrics of the training user group and classification performance.
In contrast, for the test user, significant correlations with classification performance were observed for many metrics.
In \dreyer{}, significant correlations were observed for all metrics except BtwTrialSpat (DeepConvNet) and class\_stability.
In \lee{}, significant correlations were observed for all metrics except class\_stability.
Furthermore, in both datasets, among the metrics that showed significant correlations, only MITT\_IntraTrialVar exhibited a positive correlation, whereas all other metrics showed a negative one.

Fig.~\ref{fig:cross-user_group-wise} shows the results for the cross-user (group-wise) analysis.
In \dreyer{}, significant correlations were observed only for BtwTrialGrpSpat-TR and BtwTrialGrpFreq-TR (DeepConvNet).
In \lee{}, significant correlations were observed only for BtwTrialGrpSpat-TR and BtwTrialGrpFreq-TR (TSLR).
Furthermore, for BtwTrialGrpFreq-TR, a positive correlation was observed in \dreyer{}, whereas a negative one was observed in \lee{}.
The detailed correlation results are provided in Supplementary Tables~1--4.

\section{Discussions}
\label{sec:discussions}

\subsection{Within-User Classification}
\label{sec:within-user}
Lower variability generally coincided with better BCI performance.
Among the six proposed metrics, all except BtwTrialSpat in \dreyer{} and WiTrialSpat in \lee{} exhibited negative correlations.
The within-trial metrics proposed here also showed negative correlations in many conditions, suggesting that stability of neural patterns within a trial may help explain BCI performance, whereas existing metrics mainly capture variability across trials~\cite{lotte_Definingquantifying_2018, rimbert_EventRelatedDesynchronization_2022}.

When focusing on the MITT metrics, different characteristics from the results of our proposed methods were found.
For example, MITT\_IntraTrialVar in \dreyer{} showed a positive correlation, whereas the WiTrial variability metrics proposed in this study showed negative correlations across all modalities.
This difference is likely attributable to the methodology of MITT\_IntraTrialVar.
In MITT\_IntraTrialVar, each trial is first divided into sub-epochs, and the covariance matrices obtained from these sub-epochs are clustered.
The data are then projected onto a space spanned by vectors connecting each cluster centroid and the mean of these centroids, and the temporal variation of the trajectory within this space is used as the metric.
Therefore, when the centroids of the clusters are located close to each other, and transitions between clusters occur within a trial, large values may be estimated even if the signals across sub-epochs exhibit relatively small variability and appear stable.
Consequently, even for participants who show stable neural patterns and high classification performance, large MITT\_IntraTrialVar values may be obtained if temporal changes in patterns exist, which may explain the positive correlation observed for MITT\_IntraTrialVar.

Furthermore, MITT\_InterTrialVar showed the strongest negative correlation among all metrics studied.
However, MITT\_InterTrialVar is defined as the within-class variance normalized by the total variance in the subspace obtained from the clusters formed by the set of trials. This total variance also reflects the between-class variance, i.e., class separability.
In other words, MITT\_InterTrialVar can also be interpreted as a metric that reflects not only variability, but also the degree of class separability in the subspace obtained by the MITT method.
That is, the higher the class separability, the smaller the value of MITT\_InterTrialVar.
This probably explains the strong negative correlation between MITT\_InterTrialVar and BCI performance observed here.

STDERD is defined as the standard deviation of the mean ERD/S amplitude across trials and thus quantifies between-trial variability.
Like STDERD, BtwTrialTemp evaluates between-trial variability, but captures variability in the ERD/S waveform pattern rather than only the mean amplitude.
BtwTrialTemp showed stronger correlations than STDERD in both datasets, suggesting that accounting for the ERD/S waveform pattern, rather than only its mean amplitude, provides a more informative measure of variability.

\subsection{Cross-User Classification}

In the cross-user (subject-wise) condition, an interesting result was observed.
The average within-user variability of the training users showed no significant correlation with BCI performance under any condition.
In contrast, the within-user variability of the test user showed significant correlations with BCI performance under many conditions.
These results suggest that, for cross-user BCIs, the variability of the user who actually operates the system may be more important than the variability of the users from the training set.
Therefore, from the perspective of variability, these findings suggest that reducing the variability of the user who operates the system may be an important target for future training approaches.

Furthermore, when focusing on the cross-user (group-wise) condition, no correlations with BCI performance were observed for BtwTrialGrpTemp, BtwTrialGrpSpat, or BtwTrialGrpFreq that are measuring variability between training users, in either dataset.
In contrast, when variability was evaluated relative to the test user, BtwTrialGrpSpat-TR showed positive correlations across all datasets and classification models. Furthermore, BtwTrialGrpFreq-TR also showed a positive correlation for DeepConvNet in \dreyer{}.
If these observed associations reflect causal relationships, they may suggest that higher classification performance could be achieved when the training data include users whose spatial or frequential characteristics differ from those of the test user. In other words, including users with more diverse characteristics in the training data could improve the model's generalization performance.

In particular, focusing on BtwTrialGrpSpat-TR, positive correlations were observed across all datasets and classification models.
Spatial variability across users may arise not only from individual differences such as brain anatomy but also from factors related to electrode placement.
For example, the type of EEG cap, the application of EEG gel, and slight misalignment when mounting the cap can all contribute to such variability.
Variability caused by these factors is difficult to completely avoid, given the design of current BCI systems and EEG acquisition setups.
On the other hand, the results of this study suggest that, if this observed association reflects a causal relationship, classification performance may improve when the training dataset includes users with a certain degree of variability.
One possible explanation is that using data with diverse spatial distributions improves the generalization ability of the model.
Furthermore, considering that spatial variability can arise from various measurement conditions as described above, it may be important to train models using data that explicitly account for such variability.

\begin{figure*}[!t]
\centerline{\includegraphics[width=\linewidth]{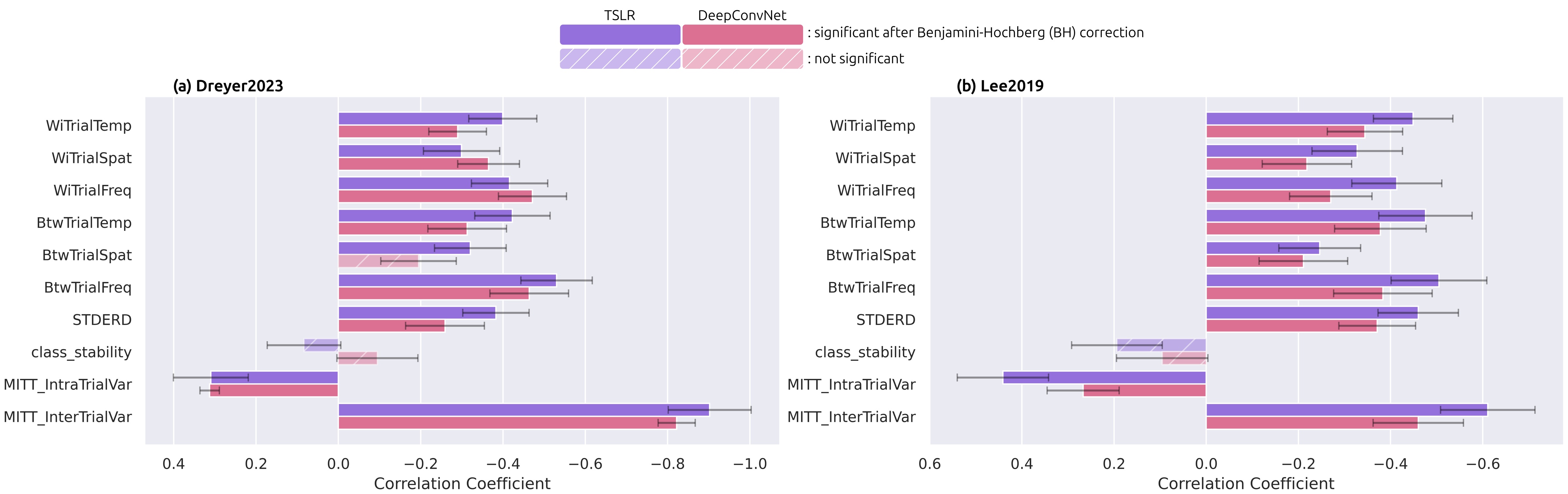}}
\caption{
Spearman correlation coefficients between cross-user (subject-wise) classification performance and each variability metric of the test user. (a) \dreyer{} and (b) \lee{}.
Purple and pink bars denote the results for TSLR and DeepConvNet, respectively.
Solid colors indicate results that are significant, whereas hatched patterns indicate non-significant correlations.
Error bars represent the standard errors.
}
\label{fig:cross-user_subject-wise}
\end{figure*}

\begin{figure*}[!t]
\centerline{\includegraphics[width=\linewidth]{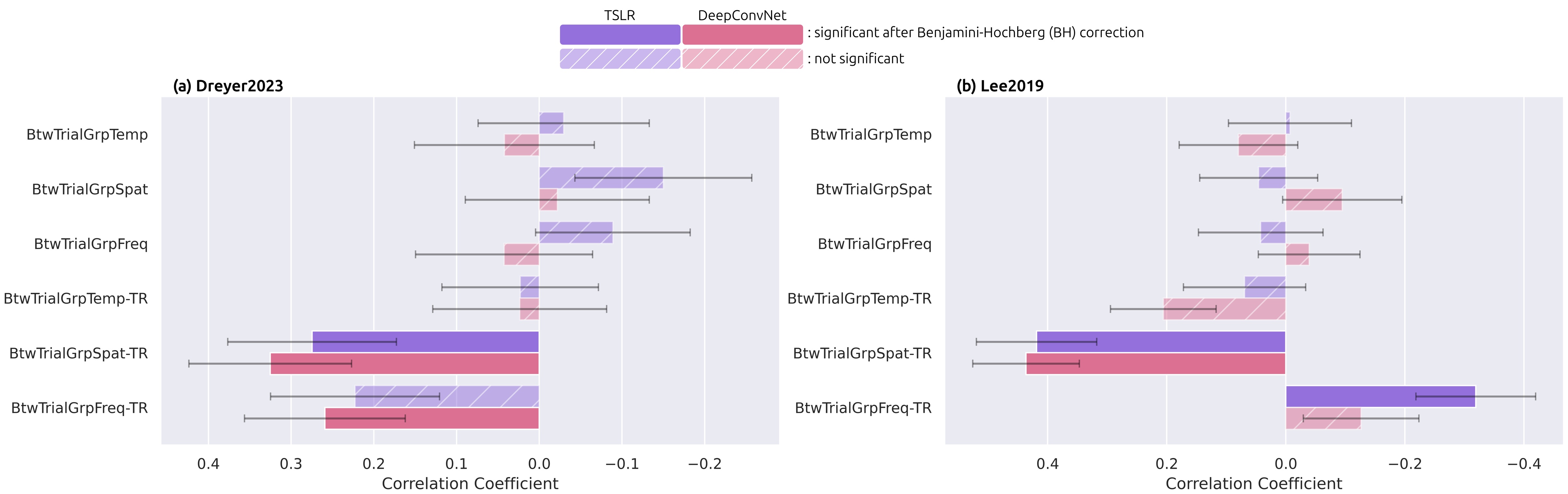}}
\caption{
Spearman correlation coefficients between cross-user (group-wise) classification performance and between-user variability metrics. (a) \dreyer{} and (b) \lee{}.
Purple and pink bars denote the results for TSLR and DeepConvNet, respectively.
Solid colors indicate results that are significant, whereas hatched patterns indicate non-significant correlations.
Error bars represent the standard errors.
}
\label{fig:cross-user_group-wise}
\end{figure*}

When focusing on the cross-user (subject-wise) analysis (Fig.~\ref{fig:cross-user_subject-wise}) based on the variability values of the test user and the cross-user (group-wise) analysis (Fig.~\ref{fig:cross-user_group-wise}), similar correlation patterns were observed across the classification models.
In particular, for all variability metrics that showed statistically significant correlations, the direction of the correlation (i.e., positive or negative) was consistent between TSLR and DeepConvNet.
These results indicate that the effect of each type of variability on classification performance is largely model-independent in terms of its direction. Despite this overall consistency, TSLR generally exhibited stronger correlation coefficients than DeepConvNet across most conditions in the cross-user (subject-wise) analysis based on the test user's variability metrics (Fig.~\ref{fig:cross-user_subject-wise}).
This suggests that the performance of TSLR may be more strongly influenced by the within-user variability of the test user than that of DeepConvNet.

\subsection{Implications of the proposed metrics}

In this study, we proposed six variability metrics, as summarized in Table.~\ref{tab:variability-metrics}.
Several metrics for quantifying EEG variability in BCIs have previously been proposed; however, to the best of our knowledge, no metric has been designed to independently evaluate variability across the three modalities of temporal, spatial, and frequency domains.
For example, STDERD measures the variability of the mean ERD amplitude across trials and can therefore be regarded as a metric that mainly reflects temporal variability.
In contrast, class\_stability and the MITT-based metrics rely on covariance matrices and can be interpreted as evaluating spatio-temporal variability.
Furthermore, as discussed in Section~\ref{sec:within-user}, MITT\_InterTrialVar may also be interpreted as measuring class separability rather than purely reflecting variability.

The metrics proposed in this study define variability based on the unified formulation given in~\eqref{eq:dispersion}, which provides a high degree of interpretability.
Moreover, by appropriately selecting the distance measure, the centroid definition, and the feature representation, the proposed framework can flexibly quantify various types of variability.
In particular, given that the understanding of EEG variability in BCI remains limited, interpretable metrics are important for advancing our understanding of variability in EEG signals.

In this study, the distance measures and centroids summarized in Table.~\ref{tab:variability-metrics} were used for the evaluation.
However, identifying the most appropriate distance measures and centroid definitions for each modality remains a topic for future work.

\subsection{Possible Applications}

It has been reported that BCI performance varies due to a variety of factors, including psychological, neurophysiological, and anatomical factors~\cite{saha_IntraIntersubject_2020a, lotte_eusipco2026}.
However, the fundamental mechanisms underlying the emergence of this variability have not yet been fully understood.
To elucidate these mechanisms, methods for quantitatively evaluating variability are essential.
By combining the metrics proposed in this study with various types of metadata and examining their relationships with variability, it may be possible to better understand the mechanisms underlying variability in BCI performance.

Variability metrics can help analyze the characteristics of classification models and datasets.
For example, in the BtwTrialSpat results shown in Fig.~\ref{fig:cross-user_subject-wise}, TSLR showed a statistically significant negative correlation, whereas DeepConvNet did not.
This suggests that DeepConvNet is relatively robust to variability in the spatial distribution of ERD/S across trials.
TSLR relies on a single spatial covariance matrix, while DeepConvNet uses multiple convolutional units that act as spatial filters, allowing it to exploit different filters depending on the variability level in the data.
This example illustrates how variability metrics can reveal how models respond to variability.
Computing these metrics for each dataset also enables comparison of variability across datasets, helping to characterize their differences.

Variability metrics can also be used as objective functions during model development, including regularization, model selection, feature selection, and preprocessing.
For instance, incorporating these metrics into the objective function enables adaptive regularization, with stronger regularization for high variability and weaker regularization for low variability.
If model behavior under variability is clarified using the approaches above, these metrics could further guide weighting of individual models in ensemble learning.
They may also support feature selection to reduce specific types of variability and the design of temporal or spatial filters.

The proposed variability metrics may also aid user training.
As shown in Fig.~\ref{fig:within-user}, many metrics were significantly negatively correlated with BCI performance, meaning lower variability tended to accompany higher performance.
Instead of only feeding back classifier output or accuracy, the variability metrics themselves could be fed back to users, enabling training strategies that directly target reducing variability.

\pagestyle{empty}
The proposed metrics may also help optimize BCI paradigms and tasks for each user. 
Because the most suitable BCI task (e.g., which mental task) and paradigm (e.g., MI, event-related potential (ERP), steady-state visual evoked potential (SSVEP)) likely differ across users
, variability could serve as one criterion for selecting tasks or paradigms.

\subsection{Applicability to Other BCI tasks}

In this study, we quantified the temporal, spatial, and frequential variability of ERD/S and investigated their relationship with MI-BCI performance. 
However, the proposed variability metrics are theoretically applicable to other BCI neural responses, such as ERP.

ERD/S data can be represented as 
$X_{\mathrm{erds}} \in \mydim{N \times C \times F \times T}$.
In contrast, ERP data can be represented as 
$X_{\mathrm{erp}} \in \mydim{N \times C \times T}$.
Therefore, similar variability analyses can be performed by extracting temporal and spatial components from $X_{\mathrm{erp}}$ as a slice of ERP.
Regarding frequency variability, frequency power needs to be computed in either case, and thus, the same analysis can also be applied to ERP data. 
However, in ERP paradigms, components such as P300, N200, and N100 occur within limited latency ranges; therefore, analyses should be performed according to each component's latency.

For SSVEP, the proposed method can be applied after converting the signals into time-frequency representations, as for ERD/S. 
Furthermore, with respect to temporal variability, SSVEP is a stimulus-locked oscillatory response~\cite{salelkar_Interactionsteadystate_2020}; therefore, if EEG segments can be extracted with sufficient precision relative to the stimulus onset, the proposed method may also be applied directly to the time-domain signals before power transformation.

\section{Conclusion}
\label{sec:conclusion}

In this study, we proposed a framework for quantifying temporal, spatial, and frequential variability of ERD/S in BCI, evaluated here for MI-BCI. By selecting appropriate features together with a centroid and distance function, the framework provides a unified and interpretable formulation for measuring multiple types of variability.

Using two datasets, we investigated the relationship between the proposed variability metrics and classification performance in within-user and cross-user settings. Significant negative correlations of approximately $-0.2$ to $-0.4$ were observed under many conditions, suggesting that lower variability tends to be associated with higher BCI performance. Moreover, the metrics revealed task- and classifier-dependent relationships, indicating that different classification models exhibit different sensitivities to specific types of variability. These findings highlight the importance of explicitly quantifying EEG variability in BCI research.

The proposed framework provides a unified way to characterize variability across modalities, facilitating the analysis of variability sources and the comparison of datasets, classifiers, and algorithms. It may also support future variability-aware training strategies, signal processing methods, and other approaches for mitigating EEG variability.
Overall, this work contributes toward the development of BCI that are more robust to variability and accessible to a wider range of users. Future work will explore alternative distance measures and validate the framework on additional BCI paradigms.

\funding{This work was funded by Inria-DFKI project NEARBY.}

\roles{Simon Kojima: Conceptualization, Formal analysis, Investigation, Methodology, Software, Validation, Visualization, Writing – original draft. Fabien Lotte: Conceptualization, Funding acquisition, Project administration, Resources, Supervision, Validation, Writing – review and editing.}

\suppdata{The supplementary material is available online. This includes additional methodological details, extended statistical results, model diagnostics, and supplementary figures supporting the findings reported in this article.}

\printbibliography

@misc{chevallier_largestEEGbased_2024,
  title = {The Largest {{EEG-based BCI}} Reproducibility Study for Open Science: The {{MOABB}} Benchmark},
  shorttitle = {The Largest {{EEG-based BCI}} Reproducibility Study for Open Science},
  author = {Chevallier, Sylvain and Carrara, Igor and Aristimunha, Bruno and Guetschel, Pierre and Sedlar, Sara and Lopes, Bruna and Velut, Sebastien and Khazem, Salim and Moreau, Thomas},
  year = 2024,
  month = apr,
  number = {arXiv:2404.15319},
  eprint = {2404.15319},
  primaryclass = {eess},
  publisher = {arXiv},
  doi = {10.48550/arXiv.2404.15319},
  urldate = {2025-08-11},
  archiveprefix = {arXiv}
}

@article{jeunet2016advances,
  title={Advances in user-training for mental-imagery-based BCI control: Psychological and cognitive factors and their neural correlates},
  author={Jeunet, Camille and N’Kaoua, Bernard and Lotte, Fabien},
  journal={Progress in brain research},
  volume={228},
  pages={3--35},
  year={2016},
  publisher={Elsevier}
}

@inproceedings{lotte_eusipco2026,
    title = {Tools, Challenges and Opportunities to Address Intra- and Inter-User Variability in {EEG-based BCIs}},
    author = {Lotte, Fabien and Kojima, Simon and Kueper, Niklas and Dreyer, Pauline and Math, Rafael and Sharma, Mansi and Roy, Raphaëlle N. and Rimbert, Sébastien and Tabie, Marc and Rekrut, Maurice},
    booktitle = {Proc. EUSIPCO},
    year = {2026},
}

@article{dreyer_largeEEG_2023,
  title = {A Large {{EEG}} Database with Users' Profile Information for Motor Imagery Brain-Computer Interface Research},
  author = {Dreyer, Pauline and Roc, Aline and Pillette, L{\'e}a and Rimbert, S{\'e}bastien and Lotte, Fabien},
  year = 2023,
  month = sep,
  journal = {Sci. Data},
  volume = {10},
  number = {1},
  pages = {580},
  publisher = {Nature Publishing Group},
  issn = {2052-4463},
  doi = {10.1038/s41597-023-02445-z},
  urldate = {2025-08-11},
  copyright = {2023 The Author(s)},
  langid = {english}
}

@article{ivanov_MultiClassIntraTrial_2026,
  title = {A {{Multi-Class Intra-Trial Trajectory Analysis Technique}} to {{Visualize}} and {{Quantify Variability}} of {{Mental Imagery EEG Signals}}},
  author = {Ivanov, Nicolas and Wong, Madeline and Chau, Tom},
  year = 2026,
  month = feb,
  journal = {Int. J. Neural Syst.},
  volume = {36},
  number = {02},
  pages = {2550075},
  publisher = {World Scientific Publishing Co.},
  issn = {0129-0657},
  doi = {10.1142/S0129065725500753},
  urldate = {2026-02-16}
}

@article{lee_EEGdataset_2019,
  title = {{{EEG}} Dataset and {{OpenBMI}} Toolbox for Three {{BCI}} Paradigms: An Investigation into {{BCI}} Illiteracy},
  shorttitle = {{{EEG}} Dataset and {{OpenBMI}} Toolbox for Three {{BCI}} Paradigms},
  author = {Lee, Min-Ho and Kwon, O-Yeon and Kim, Yong-Jeong and Kim, Hong-Kyung and Lee, Young-Eun and Williamson, John and Fazli, Siamac and Lee, Seong-Whan},
  year = 2019,
  month = may,
  journal = {GigaScience},
  volume = {8},
  number = {5},
  pages = {giz002},
  issn = {2047-217X},
  doi = {10.1093/gigascience/giz002},
  urldate = {2026-02-04}
}

@article{lotte_Definingquantifying_2018,
  title = {Defining and Quantifying Users' Mental Imagery-Based {{BCI}} Skills: A First Step},
  shorttitle = {Defining and Quantifying Users' Mental Imagery-Based {{BCI}} Skills},
  author = {Lotte, Fabien and Jeunet, Camille},
  year = 2018,
  month = jun,
  journal = {J. Neur. Eng.},
  volume = {15},
  number = {4},
  pages = {046030},
  publisher = {IOP Publishing},
  issn = {1741-2552},
  doi = {10.1088/1741-2552/aac577},
  urldate = {2025-08-11},
  langid = {english}
}

@article{pfurtscheller_EventrelatedEEG_1999,
  title = {Event-Related {{EEG}}/{{MEG}} Synchronization and Desynchronization: Basic Principles},
  shorttitle = {Event-Related {{EEG}}/{{MEG}} Synchronization and Desynchronization},
  author = {Pfurtscheller, G. and {Lopes da Silva}, F. H.},
  year = 1999,
  month = nov,
  journal = {Clin. Neurophysiol.},
  volume = {110},
  number = {11},
  pages = {1842--1857},
  issn = {1388-2457},
  doi = {10.1016/S1388-2457(99)00141-8},
  urldate = {2025-08-11}
}

@inproceedings{rimbert_EventRelatedDesynchronization_2022,
  title = {Is {{Event-Related Desynchronization}} Variability Correlated with {{BCI}} Performance?},
  booktitle = {Proc. IEEE MetroXRAINE},
  author = {Rimbert, S{\'e}bastien and Trocellier, David and Lotte, Fabien},
  year = 2022,
  month = oct,
  pages = {163--168},
  doi = {10.1109/MetroXRAINE54828.2022.9967551},
  urldate = {2025-08-11}
}

@article{sakoe_Dynamicprogramming_1978,
  title = {Dynamic Programming Algorithm Optimization for Spoken Word Recognition},
  author = {Sakoe, H. and Chiba, S.},
  year = 1978,
  month = feb,
  journal = {IEEE Trans. Acoust. Speech Signal Process.},
  volume = {26},
  number = {1},
  pages = {43--49},
  issn = {0096-3518},
  doi = {10.1109/TASSP.1978.1163055},
  urldate = {2025-08-11}
}

@article{slepian_Prolatespheroidal_1978,
  title = {Prolate Spheroidal Wave Functions, Fourier Analysis, and Uncertainty --- {{V}}: The Discrete Case},
  shorttitle = {Prolate Spheroidal Wave Functions, Fourier Analysis, and Uncertainty --- {{V}}},
  author = {Slepian, D.},
  year = 1978,
  month = may,
  journal = {Bell Syst. Tech. J.},
  volume = {57},
  number = {5},
  pages = {1371--1430},
  issn = {0005-8580},
  doi = {10.1002/j.1538-7305.1978.tb02104.x},
  urldate = {2026-02-10}
}

@article{rodrigues_RiemannianProcrustes_2019,
  title = {Riemannian {{Procrustes Analysis}}: {{Transfer Learning}} for {{Brain}}--{{Computer Interfaces}}},
  shorttitle = {Riemannian {{Procrustes Analysis}}},
  author = {Rodrigues, Pedro Luiz Coelho and Jutten, Christian and Congedo, Marco},
  year = 2019,
  month = aug,
  journal = {IEEE Trans. Biomed. Eng.},
  volume = {66},
  number = {8},
  pages = {2390--2401},
  issn = {1558-2531},
  doi = {10.1109/TBME.2018.2889705},
  urldate = {2025-08-11}
}

@article{saha_IntraIntersubject_2020a,
  title = {Intra- and {{Inter-subject Variability}} in {{EEG-Based Sensorimotor Brain Computer Interface}}: {{A Review}}},
  shorttitle = {Intra- and {{Inter-subject Variability}} in {{EEG-Based Sensorimotor Brain Computer Interface}}},
  author = {Saha, Simanto and Baumert, Mathias},
  year = 2020,
  month = jan,
  journal = {Front. Comput. Neurosci.},
  volume = {13},
  publisher = {Frontiers},
  issn = {1662-5188},
  doi = {10.3389/fncom.2019.00087},
  urldate = {2026-03-16},
  langid = {english}
}

@article{salelkar_Interactionsteadystate_2020,
  title = {Interaction between Steady-State Visually Evoked Potentials at Nearby Flicker Frequencies},
  author = {Salelkar, Siddhesh and Ray, Supratim},
  year = 2020,
  month = mar,
  journal = {Sci. Rep.},
  volume = {10},
  number = {1},
  pages = {5344},
  publisher = {Nature Publishing Group},
  issn = {2045-2322},
  doi = {10.1038/s41598-020-62180-y},
  urldate = {2026-03-16},
  copyright = {2020 The Author(s)},
  langid = {english}
}

@article{wolpaw_Braincomputerinterface_2000b,
  title = {Brain-Computer Interface Technology: A Review of the First International Meeting},
  shorttitle = {Brain-Computer Interface Technology},
  author = {Wolpaw, J.R. and Birbaumer, N. and Heetderks, W.J. and McFarland, D.J. and Peckham, P.H. and Schalk, G. and Donchin, E. and Quatrano, L.A. and Robinson, C.J. and Vaughan, T.M.},
  year = 2000,
  month = jun,
  journal = {IEEE Trans. Rehabil. Eng.},
  volume = {8},
  number = {2},
  pages = {164--173},
  issn = {1558-0024},
  doi = {10.1109/TRE.2000.847807},
  urldate = {2026-03-16}
}

@article{padfield_EEGBasedBrainComputer_2019,
  title = {{{EEG-Based Brain-Computer Interfaces Using Motor-Imagery}}: {{Techniques}} and {{Challenges}}},
  shorttitle = {{{EEG-Based Brain-Computer Interfaces Using Motor-Imagery}}},
  author = {Padfield, Natasha and Zabalza, Jaime and Zhao, Huimin and Masero, Valentin and Ren, Jinchang},
  year = 2019,
  month = mar,
  journal = {Sensors},
  volume = {19},
  number = {6},
  publisher = {Multidisciplinary Digital Publishing Institute},
  issn = {1424-8220},
  doi = {10.3390/s19061423},
  urldate = {2026-03-16},
  copyright = {http://creativecommons.org/licenses/by/3.0/},
  langid = {english}
}

@article{vidaurre_CureBCI_2010a,
  title = {Towards a {{Cure}} for {{BCI Illiteracy}}},
  author = {Vidaurre, Carmen and Blankertz, Benjamin},
  year = 2010,
  month = jun,
  journal = {Brain Topogr.},
  volume = {23},
  number = {2},
  pages = {194--198},
  issn = {1573-6792},
  doi = {10.1007/s10548-009-0121-6},
  urldate = {2026-03-16},
  langid = {english}
}

@article{hammer_Psychologicalpredictors_2012,
  title = {Psychological Predictors of {{SMR-BCI}} Performance},
  author = {Hammer, Eva Maria and Halder, Sebastian and Blankertz, Benjamin and Sannelli, Claudia and Dickhaus, Thorsten and Kleih, Sonja and M{\"u}ller, Klaus-Robert and K{\"u}bler, Andrea},
  year = 2012,
  month = jan,
  journal = {Biol. Psychol.},
  volume = {89},
  number = {1},
  pages = {80--86},
  issn = {0301-0511},
  doi = {10.1016/j.biopsycho.2011.09.006},
  urldate = {2026-03-16}
}

@article{vongroll_Largescale_2024b,
  title = {Large Scale Investigation of the Effect of Gender on Mu Rhythm Suppression in Motor Imagery Brain-Computer Interfaces},
  author = {{von Groll}, Valentina Gamboa and Leeuwis, Nikki and Rimbert, S{\'e}bastien and Roc, Aline and Pillette, L{\'e}a and Lotte, Fabien and Alimardani, Maryam},
  year = 2024,
  month = jul,
  journal = {Brain-Comput. Interfaces},
  volume = {11},
  number = {3},
  pages = {87--97},
  publisher = {Taylor \& Francis},
  issn = {2326-263X},
  doi = {10.1080/2326263X.2024.2345449},
  urldate = {2026-03-16}
}

@article{zhang_Subjectinefficiency_2020,
  title = {Subject Inefficiency Phenomenon of Motor Imagery Brain-Computer Interface: {{Influence}} Factors and Potential Solutions},
  shorttitle = {Subject Inefficiency Phenomenon of Motor Imagery Brain-Computer Interface},
  author = {Zhang, Rui and Li, Fali and Zhang, Tao and Yao, Dezhong and Xu, Peng},
  year = 2020,
  month = sep,
  journal = {Brain Sci. Adv.},
  volume = {6},
  number = {3},
  pages = {224--241},
  publisher = {SAGE Publications Ltd},
  issn = {2096-5958},
  doi = {10.26599/BSA.2020.9050021},
  urldate = {2026-03-16},
  langid = {english}
}

@article{lotte_reviewclassification_2018,
  title = {A Review of Classification Algorithms for {{EEG-based}} Brain--Computer Interfaces: A 10 Year Update},
  shorttitle = {A Review of Classification Algorithms for {{EEG-based}} Brain--Computer Interfaces},
  author = {Lotte, F and Bougrain, L and Cichocki, A and Clerc, M and Congedo, M and Rakotomamonjy, A and Yger, F},
  year = 2018,
  month = apr,
  journal = {J. Neur. Eng.},
  volume = {15},
  number = {3},
  pages = {031005},
  publisher = {IOP Publishing},
  issn = {1741-2552},
  doi = {10.1088/1741-2552/aab2f2},
  urldate = {2026-03-16},
  langid = {english}
}

@article{wu_TransferLearning_2022,
  title = {Transfer {{Learning}} for {{EEG-Based Brain}}--{{Computer Interfaces}}: {{A Review}} of {{Progress Made Since}} 2016},
  shorttitle = {Transfer {{Learning}} for {{EEG-Based Brain}}--{{Computer Interfaces}}},
  author = {Wu, Dongrui and Xu, Yifan and Lu, Bao-Liang},
  year = 2022,
  month = mar,
  journal = {IEEE Trans. Cogn. Dev. Syst.},
  volume = {14},
  number = {1},
  pages = {4--19},
  issn = {2379-8939},
  doi = {10.1109/TCDS.2020.3007453},
  urldate = {2026-03-16}
}

@article{he_TransferLearning_2020,
  title = {Transfer {{Learning}} for {{Brain}}--{{Computer Interfaces}}: {{A Euclidean Space Data Alignment Approach}}},
  shorttitle = {Transfer {{Learning}} for {{Brain}}--{{Computer Interfaces}}},
  author = {He, He and Wu, Dongrui},
  year = 2020,
  month = feb,
  journal = {IEEE Trans. Biomed. Eng.},
  volume = {67},
  number = {2},
  pages = {399--410},
  issn = {1558-2531},
  doi = {10.1109/TBME.2019.2913914},
  urldate = {2026-03-16}
}

@article{schirrmeister_Deeplearning_2017,
  title = {Deep Learning with Convolutional Neural Networks for {{EEG}} Decoding and Visualization},
  author = {Schirrmeister, Robin Tibor and Springenberg, Jost Tobias and Fiederer, Lukas Dominique Josef and Glasstetter, Martin and Eggensperger, Katharina and Tangermann, Michael and Hutter, Frank and Burgard, Wolfram and Ball, Tonio},
  year = 2017,
  journal = {Hum. Brain Mapp.},
  volume = {38},
  number = {11},
  pages = {5391--5420},
  issn = {1097-0193},
  doi = {10.1002/hbm.23730},
  urldate = {2026-03-17},
  copyright = {\copyright{} 2017 The Authors Human Brain Mapping Published by Wiley Periodicals, Inc.},
  langid = {english}
}

@article{yger_RiemannianApproaches_2017b,
  title = {Riemannian {{Approaches}} in {{Brain-Computer Interfaces}}: {{A Review}}},
  shorttitle = {Riemannian {{Approaches}} in {{Brain-Computer Interfaces}}},
  author = {Yger, Florian and Berar, Maxime and Lotte, Fabien},
  year = 2017,
  month = oct,
  journal = {IEEE Trans. Neural Syst. Rehabil. Eng.},
  volume = {25},
  number = {10},
  pages = {1753--1762},
  issn = {1558-0210},
  doi = {10.1109/TNSRE.2016.2627016},
  urldate = {2026-03-17}
}

@article{ledoit_wellconditionedestimator_2004,
  title = {A Well-Conditioned Estimator for Large-Dimensional Covariance Matrices},
  author = {Ledoit, Olivier and Wolf, Michael},
  year = 2004,
  month = feb,
  journal = {J. Multivar. Anal.},
  volume = {88},
  number = {2},
  pages = {365--411},
  issn = {0047-259X},
  doi = {10.1016/S0047-259X(03)00096-4},
  urldate = {2026-02-06}
}

@software{pyriemann,
author = {Barachant, Alexandre and Barthélemy, Quentin and King, Jean-Rémi and Gramfort, Alexandre and Chevallier, Sylvain and Rodrigues, Pedro L. C. and Olivetti, Emanuele and Goncharenko, Vladislav and Wagner vom Berg, Gabriel and Reguig, Ghiles and Lebeurrier, Arthur and Bjäreholt, Erik and Yamamoto, Maria Sayu and Clisson, Pierre and Corsi, Marie-Constance and Carrara, Igor and Mellot, Apolline and Junqueira Lopes, Bruna and Gaisford, Brent and Mian, Ammar and Andreev, Anton and Cattan, Gregoire and Aristimunha, Bruno},
doi = {10.5281/zenodo.593816},
month = {1},
title = {pyRiemann},
url = {https://doi.org/10.5281/zenodo.593816},
year = {2026}
}

@misc{aristimunha_Braindecodetoolbox_2025,
  title = {Braindecode: Toolbox for Decoding Raw Electrophysiological Brain Data with Deep Learning Models},
  shorttitle = {Braindecode},
  author = {Aristimunha, Bruno and Guetshel, Pierre and Wimpff, Martin and Gemein, Lukas and Rommel, Cedric and Banville, Hubert and Sliwowski, Maciej and Wilson, Daniel and Brandt, Simon and Gnassounou, Th{\'e}o and Paillard, Joseph and Junqueira Lopes, Bruna and Sedlar, Sara and Moreau, Thomas and Chevallier, Sylvain and Gramfort, Alexandre and Schirrmeister, Robin Tibor},
  year = 2025,
  month = dec,
  doi = {10.5281/zenodo.17699192},
  urldate = {2026-03-17},
  howpublished = {Zenodo}
}

@article{benjamini_ControllingFalse_1995,
  title = {Controlling the {{False Discovery Rate}}: {{A Practical}} and {{Powerful Approach}} to {{Multiple Testing}}},
  shorttitle = {Controlling the {{False Discovery Rate}}},
  author = {Benjamini, Yoav and Hochberg, Yosef},
  year = 1995,
  journal = {J. R. Stat. Soc. B},
  volume = {57},
  number = {1},
  pages = {289--300},
  issn = {2517-6161},
  doi = {10.1111/j.2517-6161.1995.tb02031.x},
  urldate = {2025-08-13},
  copyright = {\copyright{} 1995 Royal Statistical Society},
  langid = {english}
}

@article{demelo_procedureminimize_2024a,
  title = {A Procedure to Minimize {{EEG}} Variability for {{BCI}} Applications},
  author = {{de Melo}, Gabriel Chaves and Castellano, Gabriela and {Forner-Cordero}, Arturo},
  year = 2024,
  month = mar,
  journal = {Biomed. Signal Process. Control.},
  volume = {89},
  pages = {105745},
  issn = {1746-8094},
  doi = {10.1016/j.bspc.2023.105745},
  urldate = {2026-03-18}
}

@article{wriessnegger_InterIntraindividual_2020,
  title = {Inter- and {{Intra-individual Variability}} in {{Brain Oscillations During Sports Motor Imagery}}},
  author = {Wriessnegger, Selina C. and {M{\"u}ller-Putz}, Gernot R. and Brunner, Clemens and Sburlea, Andreea I.},
  year = 2020,
  month = oct,
  journal = {Front. Hum. Neurosci.},
  volume = {14},
  publisher = {Frontiers},
  issn = {1662-5161},
  doi = {10.3389/fnhum.2020.576241},
  urldate = {2026-03-18},
  langid = {english}
}

@article{pymanopt,
    author = {James Townsend and Niklas Koep and Sebastian Weichwald},
    journal = {J. Mach. Learn. Res.},
    number = {137},
    pages = {1–5},
    title = {Pymanopt: A Python Toolbox for Optimization on Manifolds using Automatic Differentiation},
    url = {http://jmlr.org/papers/v17/16-177.html},
    volume = {17},
    year = {2016}
}

\end{document}


\mytitle{Supplementary Materials}

\section{Supplementary Figures and Tables}

\begin{figure*}[!h]
\includegraphics[width=\linewidth]{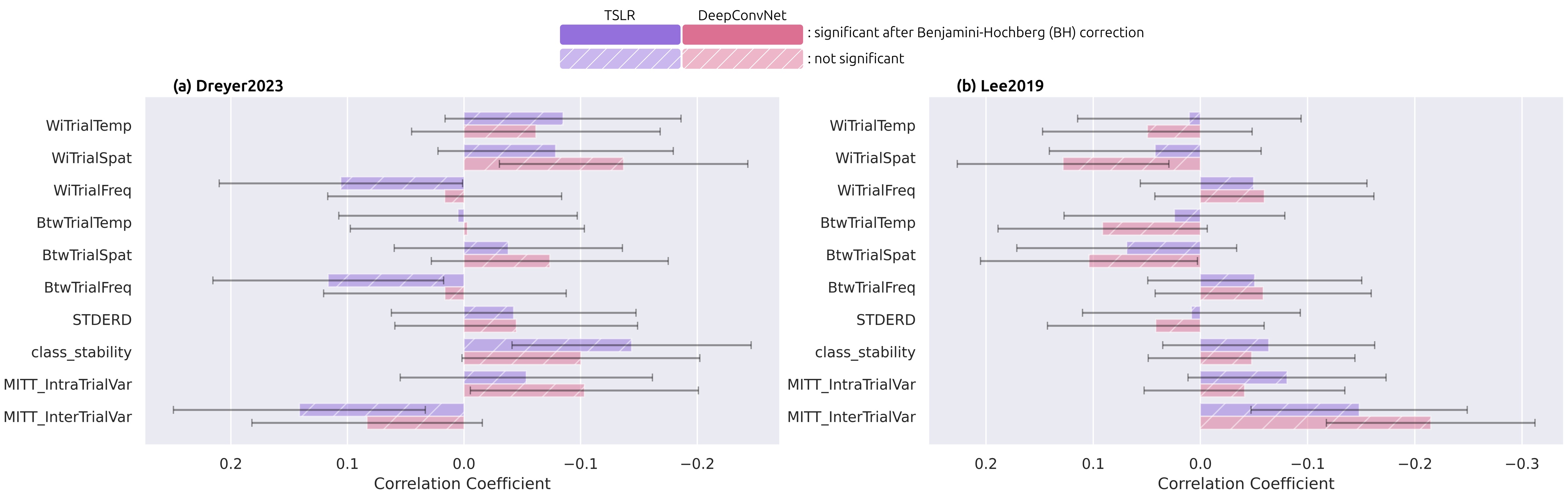}
\caption{
Spearman correlation coefficients between cross-user (subject-wise) classification performance and the mean of each variability metric across training users. (a) \dreyer{} and (b) \lee{}.
Purple and pink bars denote the results for TSLR and DeepConvNet, respectively.
Solid colors indicate results that are significant after Benjamini-Hochberg multiple comparison correction, whereas hatched patterns indicate non-significant correlations.
Error bars represent the standard errors.
}
\label{fig:cross-user_subject-wise_train}
\end{figure*}

\begin{table}[!h]
\centering
\caption{
Spearman correlation coefficients (R) between each variability metric and classification performance in within-user classification. Corresponding $p$-values were corrected using the Benjamini-Hochberg method.
SE denotes standard error.
}
\label{tab:within-user-table}
\begin{tabular}{l|l|r|r|r}
\hline
Dataset & Metrics & R & SE & $p$-value \\
\hline \hline
\multirow{10}{*}{\shortstack{\texttt{Dreyer}\\\texttt{2023}}}
& WiTrialTemp & $\mathbf{-0.32}$ & $\mathbf{0.10}$ & $\mathbf{8.94 \times 10^{-3}}$ \\
& WiTrialSpat & $\mathbf{-0.28}$ & $\mathbf{0.11}$ & $\mathbf{1.68 \times 10^{-2}}$ \\
& WiTrialFreq & $\mathbf{-0.35}$ & $\mathbf{0.10}$ & $\mathbf{7.17 \times 10^{-3}}$ \\
& BtwTrialTemp & $\mathbf{-0.3}$ & $\mathbf{0.11}$ & $\mathbf{1.19 \times 10^{-2}}$ \\
& BtwTrialSpat & $-0.11$ & $0.11$ & $3.64 \times 10^{-1}$ \\
& BtwTrialFreq & $\mathbf{-0.33}$ & $\mathbf{0.10}$ & $\mathbf{7.17 \times 10^{-3}}$ \\
& STDERD & $\mathbf{-0.26}$ & $\mathbf{0.11}$ & $\mathbf{2.24 \times 10^{-2}}$ \\
& class\_stability & $0.07$ & $0.11$ & $5.35 \times 10^{-1}$ \\
& Mitt\_IntraTrialVar & $\mathbf{0.28}$ & $\mathbf{0.11}$ & $\mathbf{1.68 \times 10^{-2}}$ \\
& Mitt\_InterTrialVar & $\mathbf{-0.84}$ & $\mathbf{0.043}$ & $\mathbf{2.34 \times 10^{-22}}$ \\

\cline{1-5}

\multirow{10}{*}{\shortstack{\texttt{Lee}\\\texttt{2019}}}
& WiTrialTemp & $\mathbf{-0.41}$ & $\mathbf{0.12}$ & $\mathbf{9.06 \times 10^{-3}}$ \\
& WiTrialSpat & $-0.19$ & $0.14$ & $2.41 \times 10^{-1}$ \\
& WiTrialFreq & $\mathbf{-0.37}$ & $\mathbf{0.14}$ & $\mathbf{1.40 \times 10^{-2}}$ \\
& BtwTrialTemp & $\mathbf{-0.41}$ & $\mathbf{0.13}$ & $\mathbf{9.06 \times 10^{-3}}$ \\
& BtwTrialSpat & $\mathbf{-0.31}$ & $\mathbf{0.14}$ & $\mathbf{3.88 \times 10^{-2}}$ \\
& BtwTrialFreq & $\mathbf{-0.4}$ & $\mathbf{0.13}$ & $\mathbf{9.37 \times 10^{-3}}$ \\
& STDERD & $\mathbf{-0.36}$ & $\mathbf{0.13}$ & $\mathbf{1.45 \times 10^{-2}}$ \\
& class\_stability & $0.16$ & $0.13$ & $3.05 \times 10^{-1}$ \\
& Mitt\_IntraTrialVar & $-0.037$ & $0.14$ & $7.95 \times 10^{-1}$ \\
& Mitt\_InterTrialVar & $\mathbf{-0.64}$ & $\mathbf{0.096}$ & $\mathbf{5.51 \times 10^{-6}}$ \\
\cline{1-5}
\end{tabular}
\end{table}

\begin{table}[h]
\centering
\caption{
Spearman correlation coefficients (R) between each variability metric of test user and classification performance in cross-user classification. Corresponding $p$-values were corrected using the Benjamini-Hochberg method.
SE denotes standard error.
}
\label{tab:within-user-table}
\begin{tabular}{l|l|l|r|r|r}
\hline
Dataset & Metrics & Model & R & SE & $p$-value \\
\hline \hline
\multirow{20}{*}{\shortstack{\texttt{Dreyer}\\\texttt{2023}}}
& \multirow{2}{*}{WiTrialTemp} & TSLR & $\mathbf{-0.4}$ & $\mathbf{0.083}$ & $\mathbf{9.59 \times 10^{-5}}$ \\
& & DeepConvNet & $\mathbf{-0.29}$ & $\mathbf{0.093}$ & $\mathbf{4.31 \times 10^{-3}}$ \\
\cline{2-6}
& \multirow{2}{*}{WiTrialSpat} & TSLR & $\mathbf{-0.3}$ & $\mathbf{0.093}$ & $\mathbf{3.31 \times 10^{-3}}$ \\
& & DeepConvNet & $\mathbf{-0.37}$ & $\mathbf{0.091}$ & $\mathbf{3.77 \times 10^{-4}}$ \\
\cline{2-6}
& \multirow{2}{*}{WiTrialFreq} & TSLR & $\mathbf{-0.42}$ & $\mathbf{0.087}$ & $\mathbf{4.77 \times 10^{-5}}$ \\
& & DeepConvNet & $\mathbf{-0.47}$ & $\mathbf{0.087}$ & $\mathbf{3.65 \times 10^{-6}}$ \\
\cline{2-6}
& \multirow{2}{*}{BtwTrialTemp} & TSLR & $\mathbf{-0.42}$ & $\mathbf{0.08}$ & $\mathbf{3.89 \times 10^{-5}}$ \\
& & DeepConvNet & $\mathbf{-0.31}$ & $\mathbf{0.09}$ & $\mathbf{2.35 \times 10^{-3}}$ \\
\cline{2-6}
& \multirow{2}{*}{BtwTrialSpat} & TSLR & $\mathbf{-0.32}$ & $\mathbf{0.091}$ & $\mathbf{2.08 \times 10^{-3}}$ \\
& & DeepConvNet & $-0.19$ & 0.10 & $5.78 \times 10^{-2}$ \\
\cline{2-6}
& \multirow{2}{*}{BtwTrialFreq} & TSLR & $\mathbf{-0.53}$ & $\mathbf{0.07}$ & $\mathbf{9.36 \times 10^{-8}}$ \\
& & DeepConvNet & $\mathbf{-0.46}$ & $\mathbf{0.075}$ & $\mathbf{4.66 \times 10^{-6}}$ \\
\cline{2-6}
& \multirow{2}{*}{STDERD} & TSLR & $\mathbf{-0.38}$ & $\mathbf{0.083}$ & $\mathbf{1.88 \times 10^{-4}}$ \\
& & DeepConvNet & $\mathbf{-0.26}$ & $\mathbf{0.096}$ & $\mathbf{1.08 \times 10^{-2}}$ \\
\cline{2-6}
& \multirow{2}{*}{class\_stability} & TSLR & $0.084$ & $0.092$ & $4.09 \times 10^{-1}$ \\
& & DeepConvNet & $-0.095$ & 0.096 & $3.67 \times 10^{-1}$ \\
\cline{2-6}
& \multirow{2}{*}{Mitt\_IntraTrialVar} & TSLR & $\mathbf{0.31}$ & $\mathbf{0.096}$ & $\mathbf{2.43 \times 10^{-3}}$ \\
& & DeepConvNet & $\mathbf{0.31}$ & $\mathbf{0.098}$ & $\mathbf{2.35 \times 10^{-3}}$ \\
\cline{2-6}
& \multirow{2}{*}{Mitt\_InterTrialVar} & TSLR & $\mathbf{-0.9}$ & $\mathbf{0.024}$ & $\mathbf{2.67 \times 10^{-36}}$ \\
& & DeepConvNet & $\mathbf{-0.82}$ & $\mathbf{0.045}$ & $\mathbf{1.04 \times 10^{-24}}$ \\

\cline{1-6}

& \multirow{2}{*}{WiTrialTemp} & TSLR & $\mathbf{-0.45}$ & $\mathbf{0.086}$ & $\mathbf{9.34 \times 10^{-6}}$ \\
& & DeepConvNet & $\mathbf{-0.34}$ & $\mathbf{0.098}$ & $\mathbf{7.52 \times 10^{-4}}$ \\
\cline{2-6}
& \multirow{2}{*}{WiTrialSpat} & TSLR & $\mathbf{-0.33}$ & $\mathbf{0.098}$ & $\mathbf{1.35 \times 10^{-3}}$ \\
& & DeepConvNet & $\mathbf{-0.22}$ & $\mathbf{0.10}$ & $\mathbf{3.40 \times 10^{-2}}$ \\
\cline{2-6}
& \multirow{2}{*}{WiTrialFreq} & TSLR & $\mathbf{-0.41}$ & $\mathbf{0.089}$ & $\mathbf{4.77 \times 10^{-5}}$ \\
& & DeepConvNet & $\mathbf{-0.27}$ & $\mathbf{0.10}$ & $\mathbf{9.34 \times 10^{-3}}$ \\
\cline{2-6}
& \multirow{2}{*}{BtwTrialTemp} & TSLR & $\mathbf{-0.48}$ & $\mathbf{0.087}$ & $\mathbf{3.75 \times 10^{-6}}$ \\
& & DeepConvNet & $\mathbf{-0.38}$ & $\mathbf{0.098}$ & $\mathbf{2.12 \times 10^{-4}}$ \\
\cline{2-6}
& \multirow{2}{*}{BtwTrialSpat} & TSLR & $\mathbf{-0.25}$ & $\mathbf{0.099}$ & $\mathbf{1.69 \times 10^{-2}}$ \\
& & DeepConvNet & $\mathbf{-0.21}$ & $\mathbf{0.10}$ & $\mathbf{3.92 \times 10^{-2}}$ \\
\cline{2-6}
& \multirow{2}{*}{BtwTrialFreq} & TSLR & $\mathbf{-0.51}$ & $\mathbf{0.082}$ & $\mathbf{8.29 \times 10^{-7}}$ \\
& & DeepConvNet & $\mathbf{-0.38}$ & $\mathbf{0.097}$ & $\mathbf{1.84 \times 10^{-4}}$ \\
\cline{2-6}
& \multirow{2}{*}{STDERD} & TSLR & $\mathbf{-0.46}$ & $\mathbf{0.09}$ & $\mathbf{5.89 \times 10^{-6}}$ \\
& & DeepConvNet & $\mathbf{-0.37}$ & $\mathbf{0.099}$ & $\mathbf{2.62 \times 10^{-4}}$ \\
\cline{2-6}
& \multirow{2}{*}{class\_stability} & TSLR & $0.19$ & $0.096$ & $5.61 \times 10^{-2}$ \\
& & DeepConvNet & $0.096$ & 0.11 & $3.44 \times 10^{-1}$ \\
\cline{2-6}
& \multirow{2}{*}{Mitt\_IntraTrialVar} & TSLR & $\mathbf{0.44}$ & $\mathbf{0.083}$ & $\mathbf{1.24 \times 10^{-5}}$ \\
& & DeepConvNet & $\mathbf{0.27}$ & $\mathbf{0.099}$ & $\mathbf{9.62 \times 10^{-3}}$ \\
\cline{2-6}
& \multirow{2}{*}{Mitt\_InterTrialVar} & TSLR & $\mathbf{-0.61}$ & $\mathbf{0.078}$ & $\mathbf{2.94 \times 10^{-10}}$ \\
& & DeepConvNet & $\mathbf{-0.46}$ & $\mathbf{0.098}$ & $\mathbf{5.89 \times 10^{-6}}$ \\
\cline{1-6}
\end{tabular}
\end{table}

\begin{table}[h]
\centering
\caption{cross-user (group-wise)
Spearman correlation coefficients (R) between cross-user classification performance and between-user variability metrics. Corresponding $p$-values were corrected using the Benjamini-Hochberg method.
SE denotes standard error.
}
\label{tab:within-user-table}
\begin{tabular}{l|l|l|r|r|r}
\hline
Dataset & Metrics & Model & R & SE & $p$-value \\
\hline \hline
\multirow{12}{*}{\shortstack{\texttt{Dreyer}\\\texttt{2023}}}
& \multirow{2}{*}{BtwTrialGrpTemp} & TSLR & $-0.029$ & $0.10$ & $8.29 \times 10^{-1}$ \\
& & DeepConvNet & $0.042$ & $0.11$ & $8.29 \times 10^{-1}$ \\
\cline{2-6}
& \multirow{2}{*}{BtwTrialGrpSpat} & TSLR & $-0.15$ & $0.093$ & $3.27 \times 10^{-1}$ \\
& & DeepConvNet & $-0.022$ & $0.095$ & $8.29 \times 10^{-1}$ \\
\cline{2-6}
& \multirow{2}{*}{BtwTrialGrpFreq} & TSLR & $-0.089$ & $0.10$ & $7.57 \times 10^{-1}$ \\
& & DeepConvNet & $0.043$ & $0.10$ & $8.29 \times 10^{-1}$ \\
\cline{2-6}
& \multirow{2}{*}{BtwTrialGrpTemp-TR} & TSLR & $0.023$ & $0.11$ & $8.29 \times 10^{-1}$ \\
& & DeepConvNet & $0.024$ & $0.11$ & $8.29 \times 10^{-1}$ \\
\cline{2-6}
& \multirow{2}{*}{BtwTrialGrpSpat-TR} & TSLR & $\mathbf{0.27}$ & $\mathbf{0.11}$ & $\mathbf{3.41 \times 10^{-2}}$ \\
& & DeepConvNet & $\mathbf{0.33}$ & $\mathbf{0.11}$ & $\mathbf{1.15 \times 10^{-2}}$ \\
\cline{2-6}
& \multirow{2}{*}{BtwTrialGrpFreq-TR} & TSLR & $0.22$ & $0.098$ & $7.78 \times 10^{-2}$ \\
& & DeepConvNet & $\mathbf{0.26}$ & $\mathbf{0.097}$ & $\mathbf{3.68 \times 10^{-2}}$ \\

\cline{1-6}

\multirow{12}{*}{\shortstack{\texttt{Lee}\\\texttt{2019}}}

& \multirow{2}{*}{BtwTrialGrpTemp} & TSLR & $-0.007$ & $0.10$ & $9.45 \times 10^{-1}$ \\
& & DeepConvNet & $0.08$ & $0.099$ & $7.39 \times 10^{-1}$ \\
\cline{2-6}
& \multirow{2}{*}{BtwTrialGrpSpat} & TSLR & $0.046$ & $0.10$ & $7.62 \times 10^{-1}$ \\
& & DeepConvNet & $-0.095$ & $0.10$ & $6.98 \times 10^{-1}$ \\
\cline{2-6}
& \multirow{2}{*}{BtwTrialGrpFreq} & TSLR & $0.042$ & $0.10$ & $7.62 \times 10^{-1}$ \\
& & DeepConvNet & $-0.039$ & $0.10$ & $7.62 \times 10^{-1}$ \\
\cline{2-6}
& \multirow{2}{*}{BtwTrialGrpTemp-TR} & TSLR & $0.069$ & $0.10$ & $7.39 \times 10^{-1}$ \\
& & DeepConvNet & $0.21$ & $0.10$ & $1.20 \times 10^{-1}$ \\
\cline{2-6}
& \multirow{2}{*}{BtwTrialGrpSpat-TR} & TSLR & $\mathbf{0.42}$ & $\mathbf{0.085}$ & $\mathbf{8.72 \times 10^{-5}}$ \\
& & DeepConvNet & $\mathbf{0.44}$ & $\mathbf{0.089}$ & $\mathbf{6.80 \times 10^{-5}}$ \\
\cline{2-6}
& \multirow{2}{*}{BtwTrialGrpFreq-TR} & TSLR & $\mathbf{-0.32}$ & $\mathbf{0.09}$ & $\mathbf{4.87 \times 10^{-3}}$ \\
& & DeepConvNet & $-0.13$ & $0.097$ & $5.03 \times 10^{-1}$ \\

\cline{1-6}
\end{tabular}
\end{table}

\begin{table}[h]
\centering
\caption{
Spearman correlation coefficients between cross-user classification performance and the mean of each variability metric across training users.
Corresponding $p$-values were corrected using the Benjamini-Hochberg method.
SE denotes standard error.
}
\label{tab:within-user-table}
\begin{tabular}{l|l|l|r|r|r}
\hline
Dataset & Metrics & Model & R & SE & $p$-value \\
\hline \hline
\multirow{20}{*}{\shortstack{\texttt{Dreyer}\\\texttt{2023}}}
& \multirow{2}{*}{WiTrialTemp} & TSLR & $-0.085$ & $0.10$ & $8.49 \times 10^{-1}$ \\
& & DeepConvNet & $-0.062$ & 0.10 & $8.85 \times 10^{-1}$ \\
\cline{2-6}
& \multirow{2}{*}{WiTrialSpat} & TSLR & $-0.079$ & $0.10$ & $8.49 \times 10^{-1}$ \\
& & DeepConvNet & $-0.14$ & 0.10 & $8.49 \times 10^{-1}$ \\
\cline{2-6}
& \multirow{2}{*}{WiTrialFreq} & TSLR & $0.11$ & $0.098$ & $8.49 \times 10^{-1}$ \\
& & DeepConvNet & $0.017$ & 0.099 & $9.68 \times 10^{-1}$ \\
\cline{2-6}
& \multirow{2}{*}{BtwTrialTemp} & TSLR & $0.0052$ & $0.10$ & $9.78 \times 10^{-1}$ \\
& & DeepConvNet & $-0.0028$ & 0.10 & $9.78 \times 10^{-1}$ \\
\cline{2-6}
& \multirow{2}{*}{BtwTrialSpat} & TSLR & $-0.038$ & $0.11$ & $8.85 \times 10^{-1}$ \\
& & DeepConvNet & $-0.074$ & 0.11 & $8.49 \times 10^{-1}$ \\
\cline{2-6}
& \multirow{2}{*}{BtwTrialFreq} & TSLR & $0.12$ & $0.11$ & $8.49 \times 10^{-1}$ \\
& & DeepConvNet & $0.016$ & 0.11 & $9.68 \times 10^{-1}$ \\
\cline{2-6}
& \multirow{2}{*}{STDERD} & TSLR & $-0.043$ & $0.10$ & $8.85 \times 10^{-1}$ \\
& & DeepConvNet & $-0.045$ & 0.10 & $8.85 \times 10^{-1}$ \\
\cline{2-6}
& \multirow{2}{*}{class\_stability} & TSLR & $-0.14$ & $0.10$ & $8.49 \times 10^{-1}$ \\
& & DeepConvNet & $-0.10$ & 0.10 & $8.49 \times 10^{-1}$ \\
\cline{2-6}
& \multirow{2}{*}{Mitt\_IntraTrialVar} & TSLR & $-0.053$ & $0.10$ & $8.85 \times 10^{-1}$ \\
& & DeepConvNet & $-0.10$ & 0.10 & $8.49 \times 10^{-1}$ \\
\cline{2-6}
& \multirow{2}{*}{Mitt\_InterTrialVar} & TSLR & $0.14$ & $0.098$ & $8.49 \times 10^{-1}$ \\
& & DeepConvNet & $0.083$ & 0.099 & $8.49 \times 10^{-1}$ \\

\cline{1-6}

\multirow{20}{*}{\shortstack{\texttt{Lee}\\\texttt{2019}}}

& \multirow{2}{*}{WiTrialTemp} & TSLR & $0.01$ & $0.10$ & $9.34 \times 10^{-1}$ \\
& & DeepConvNet & $0.049$ & 0.099 & $8.05 \times 10^{-1}$ \\
\cline{2-6}
& \multirow{2}{*}{WiTrialSpat} & TSLR & $0.042$ & $0.11$ & $8.05 \times 10^{-1}$ \\
& & DeepConvNet & $0.13$ & 0.10 & $8.05 \times 10^{-1}$ \\
\cline{2-6}
& \multirow{2}{*}{WiTrialFreq} & TSLR & $-0.05$ & $0.10$ & $8.05 \times 10^{-1}$ \\
& & DeepConvNet & $-0.06$ & 0.10 & $8.05 \times 10^{-1}$ \\
\cline{2-6}
& \multirow{2}{*}{BtwTrialTemp} & TSLR & $0.024$ & $0.10$ & $9.01 \times 10^{-1}$ \\
& & DeepConvNet & $0.091$ & 0.099 & $8.05 \times 10^{-1}$ \\
\cline{2-6}
& \multirow{2}{*}{BtwTrialSpat} & TSLR & $0.069$ & $0.092$ & $8.05 \times 10^{-1}$ \\
& & DeepConvNet & $0.10$ & 0.10 & $8.05 \times 10^{-1}$ \\
\cline{2-6}
& \multirow{2}{*}{BtwTrialFreq} & TSLR & $-0.051$ & $0.098$ & $8.05 \times 10^{-1}$ \\
& & DeepConvNet & $-0.059$ & 0.099 & $8.05 \times 10^{-1}$ \\
\cline{2-6}
& \multirow{2}{*}{STDERD} & TSLR & $0.0084$ & $0.10$ & $9.34 \times 10^{-1}$ \\
& & DeepConvNet & $0.042$ & 0.098 & $8.05 \times 10^{-1}$ \\
\cline{2-6}
& \multirow{2}{*}{class\_stability} & TSLR & $-0.064$ & $0.10$ & $8.05 \times 10^{-1}$ \\
& & DeepConvNet & $-0.048$ & 0.10 & $8.05 \times 10^{-1}$ \\
\cline{2-6}
& \multirow{2}{*}{Mitt\_IntraTrialVar} & TSLR & $-0.081$ & $0.10$ & $8.05 \times 10^{-1}$ \\
& & DeepConvNet & $-0.041$ & 0.096 & $8.05 \times 10^{-1}$ \\
\cline{2-6}
& \multirow{2}{*}{Mitt\_InterTrialVar} & TSLR & $-0.15$ & $0.094$ & $8.05 \times 10^{-1}$ \\
& & DeepConvNet & $-0.21$ & 0.097 & $6.37 \times 10^{-1}$ \\

\cline{1-6}
\end{tabular}
\end{table}